\documentclass[aps,prb,preprint]{revtex4-1}
\usepackage[latin1]{inputenc}
\usepackage{amsmath}
\usepackage{amsfonts}
\usepackage{amssymb}
\usepackage{bm}
\usepackage{natbib}
\usepackage{graphicx}

\begin{document}
\title{Strain driven migration of In during the growth of InAs/GaAs quantum posts}
\date{\today}
\author{D. Alonso-\'Alvarez}
\email{d.alonso-alvarez@imperial.ac.uk}
\altaffiliation[Present address: ]{Experimental Solid State Physics, Department of Physics, Blackett Laboratory, Imperial College, London SW7 2AZ, United Kingdom}
\author{B. Al\'en}
\author{J. M. Ripalda}
\email[Corresponding Author: ]{ripalda@imm.cnm.csic.es}
\author{A. Rivera}
\author{A. G. Taboada}
\author{J. M. Llorens}
\author{Y. Gonz\'alez}
\author{L. Gonz\'alez}
\author{F. Briones}
\affiliation{IMM-Instituto de Microelectr\'onica de Madrid (CNM-CSIC), Isaac Newton 8, 28760 Tres Cantos, Spain}


\begin{abstract}
Using the mechano-optical stress sensor technique, we observe a counter-intuitive reduction of the compressive stress when InAs is deposited on GaAs (001) during growth of quantum posts. Through modelling of the strain fields, we find that such anomalous behaviour can be related to the strain-driven detachment of In atoms from the crystal and their surface diffusion towards the self-assembled nanostructures.
\end{abstract}
\maketitle

The segregation and surface diffusion of In atoms during growth of InAs/GaAs quantum nanostructures has been a matter of debate during the last decade. Garc\'ia~\emph{et al.} suggested that the existence of a stress free, liquid-like, In layer at the crystal surface could explain the evolution of the accumulated stress measured \textit{in-situ} during the growth of InAs self-assembled quantum dots (QDs) on GaAs (001).\cite{garcia2000} Bottomley made a thermodynamical analysis of the InAs/GaAs (001) interface and extended such interpretation postulating the simultaneous formation of In and InAs liquid-like phases.\cite{bottomley2002} To explain the QD size evolution during growth, the strain driven surface diffusion of In and Ga from the WL to the QDs and \emph{vice versa} was postulated in different theoretical works.\cite{Tersoff1998, Koduvely1999} Joyce~\emph{et al.} made a study varying the InAs growth rate which suggested that the strain field gradients assist the migration of In atoms from the wetting  layer (WL) to the QDs.\cite{joyce2000} Additionally, Bottomley suggested that this In transfer from the WL might be favoured by the existence of liquid phases at the surface. A direct experimental demonstration for such mechanisms could not be given at that moment, though.

In this letter, we  present experimental evidence of these migration processes during the fabrication of InAs/GaAs quantum posts (QPs), a heteroepitaxial strained system where the strain field is the dominant factor in the growth kinetics. 

Despite the debate around certain aspects related with their growth, InAs/GaAs self-assembled nanostructures are among the best known heteroepitaxial systems.\cite{wang2008} One of the latest developments has been the fabrication of InAs/GaAs QPs\cite{he2007, li2007, krenner2008} These are elongated,  vertical nanostructures which allow for exceedingly tunable exciton radiative lifetimes, estimated from few nanoseconds to tenths of milliseconds at low temperatures, when embedded in a vertical field effect device.\cite{he2007} Such feature is of interest for quantum memories or highly non-linear electro-optical devices.\cite{krenner2008} Moreover, since their light polarization properties can be tailored controlling their height, they are also interesting for polarization sensitive applications like semiconductor optical amplifiers.\cite{li2007, sugawara2004} To this respect, we have recently reported how the introduction of phosphorus in the GaAs barrier partially balances the compressive strain of the QP enabling the fabrication of even larger nanostructures (120 nm high).\cite{alonso-alvarez2011}

InAs QPs are synthesized by growing a short period InAs/GaAs superlattice on top of a QD seed layer and therefore are related in nature to vertically coupled quantum dots.\cite{he2004, alonso-alvarez2011-2} In such context, since QPs formation heavily relies on a strain driven process and an efficient surface diffusion of In adatoms towards energetically favourable sites, they are an exemplary system where to study alloying, segregation and surface diffusion effects. In the following, we present accumulated stress measurements recorded \emph{in-situ} and in real time while growing InAs/GaAs QPs by means of the Mechano-Optical Stress Sensor technique (MOSS).\cite{garcia2000}\footnote{See Supplemental Material at [URL will be inserted by publisher] for more information about the implementation and theory of the MOSS technique.} 

\begin{figure}
	\centering
	\includegraphics{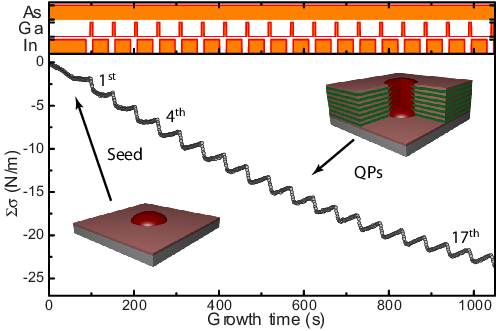}
	\caption{Experimental MOSS curve of QPs. A more negative value of $\Sigma\sigma$ indicates an increase of the compressive stress. The opening (orange) - closing (white) sequence of the effusion cells is shown on top.}
	\label{fig:Fig1}
\end{figure}

The QPs studied in this work have been grown using solid source molecular beam epitaxy (MBE) on GaAs (001) substrates with a thickness of 100 $\mu$m. Substrate temperature and As beam equivalent pressure (BEP) are kept at 510$\,^{\circ}\text{C}$ and 1.5$\times$10$^{-6}$ mbar, respectively. InAs and GaAs growth rates are 0.02 ML/s and 0.5 ML/s, respectively. Under these conditions, the QD seed layer was fabricated with the deposition of 2 ML (100 s) of InAs capped with 3 ML (6 s) of GaAs. Then, QPs were formed cycling 20 periods of 0.7 ML (35 s) of InAs and 3 ML (6 s) of GaAs to form the quantum posts while monitoring the accumulated stress ($\Sigma\sigma$). This sequence is similar to that used by Li \textit{et al}.\cite{li2007} 

We have recently shown that the monitoring of $\Sigma\sigma$ in real time is a valuable tool to optimize the growth of vertical stacks of quantum dots using different strain compensation methods.\cite{alonso-alvarez2011-2} Thanks to the high sensitivity of our setup (better than 0.1 N/m), each stage of the growth sequence produces a sizeable change of the accumulated stress curve. These changes respond to variations in the growth kinetics which are hardly detectable by other \textit{in-situ} or \textit{ex-situ} techniques.

Fig.~\ref{fig:Fig1} shows the MOSS curve recorded during growth of the QPs. It begins with the deposition of the QD seed layer and thus reproduces the accumulated stress evolution known for InAs QDs.\cite{garcia2000} The compressive stress increases ($\Sigma\sigma$ becomes more negative) following a two-step-sequence. The first step occurs during the In deposition due to the incorporation of In atoms to the crystal. The second step takes place during the Ga deposition due to the incorporation to the crystal of the liquid like In-phase that remains at the surface. Depending on the amount of deposited In and on the temperature, the composition of the resulting InGaAs layer will vary and hence the total accumulated stress.

\begin{figure}
	\centering
	\includegraphics{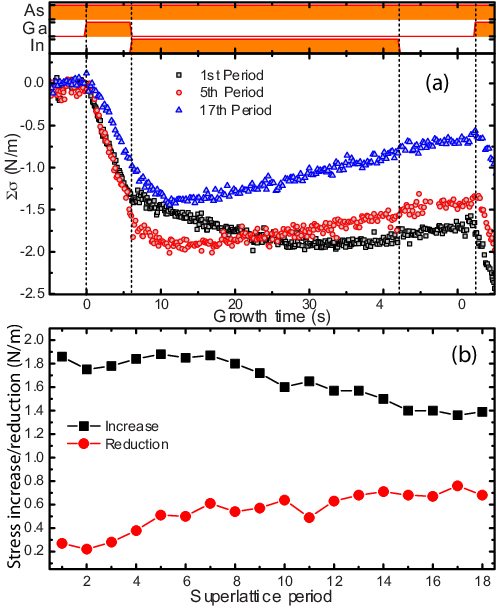}
	\caption{(a) Detail of the stress evolution of three periods of the QPs growth sequence overlapped for comparison. (b) Partial increase/reduction of the accumulated stress observed in each period.}
	\label{fig:Fig2}
\end{figure} 

At this point, the 20 period sequence used to fabricate the QPs begins. The induced stress is better appreciated in Fig. 2a where we show a detail of the $\Sigma\sigma$ curve recorded during three different periods.  The opening/closing times of the As, Ga and In effusion cells are indicated. During the first periods, the accumulated stress evolves following basically the same two steps already discussed for the seed layer. Nevertheless, it is evident that the slope of the $\Sigma\sigma$ curve changes sign (becoming positive) during the As pause, indicating an overall reduction of the compressive stress. From the 4 th period and onwards this reduction of the compressive stress occurs also while In is being deposited, which is unforeseen, though, since In incorporation is expected to steadily increase the accumulated stress due to its larger covalent radius in comparison with Ga. As it can be seen in Figure 2b, with each new period in the sequence, this reduction of the compressive stress caused by In deposition is more pronounced (red circles). At the same time, the deposition of Ga also causes less compressive stress (black squares).

To understand this anomalous behaviour we shall compare our results with accumulated stress experiments performed in other strained heteroepitaxial systems.  InAs/GaAs QDs are grown in the Stransky-Krastanow growth mode and, therefore, a reduction in the slope of the $\Sigma\sigma$ curve is expected when the two dimensional-three dimensional (2D/3D) strain relaxation occurs.\cite{garcia2000} However, a reversal of the slope sign as observed here is absent for InAs/GaAs QDs, either during the pause under As flux or during the GaAs capping. On the contrary, an slope sign change was observed by Gonz\'alez \textit{et al.} during the fabrication of InAs/InP quantum wires.\cite{Gonzalez:2004ev} The change took place just after finishing the InAs deposition, during the As pause.  Before that, the deposition of InAs increased the compressive stress linearly. They attributed this reduction to the surface relaxation and mass redistribution that led to the formation of the nanostructures. Such mass transfer from the WL - and even from the underlying material - to the nanostructures could be later assessed during the fabrication of stacked InAs/InP quantum wires.\cite{fuster2004} The mass transfer in the upper layers of the stack was directly observed by transmission electron microscopy and correlated to the strain fields created by the buried nanostructures. Finally, Silveira \textit{et al.}  also observed a  reduction of the compressive stress during growth of GaSb/GaAs QDs.\cite{silveira2001} The change of sign occurred at the critical thickness for QDs nucleation but before the end of the GaSb deposition.   In their analysis, they concluded that this reduction was also a consequence of a large mass transfer from the strained 2D layer to form the 3D nanostructures.

\begin{figure}
	\centering
	\includegraphics{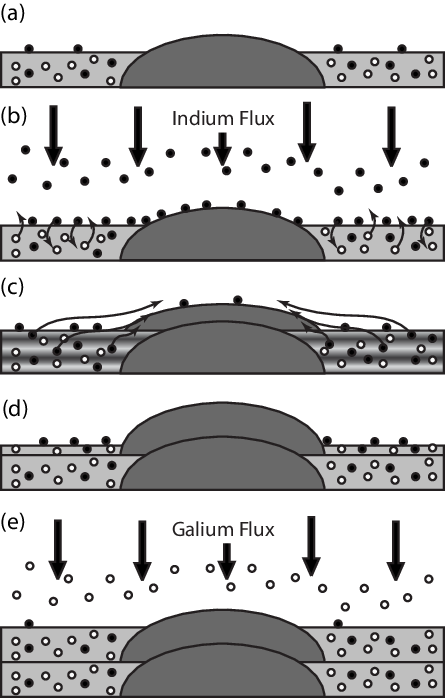}
	\caption{Sequence of steps describing the growth of a single period of a QP: a) Surface before the InAs deposition, with the top of the QPs 
half buried. b) During InAs deposition, the surface saturates with In atoms and there is a strong intermixing with Ga of the last layers. c) Due to the augmented stress the surface partially melts and ,due to the strain gradients, In atoms migrate towards the QPs. d) A fraction of the In atoms remains on the surface in the liquid-like phase. e) These atoms are incorporated to the crystal during the deposition of the GaAs capping restoring the growth front to state a). $\circ$ Ga atoms; $\bullet$ In atoms.}
	\label{fig:Fig3}
\end{figure} 

The compressive stress reduction during In deposition can only be explained by relocation of the In atoms into a configuration that makes a smaller contribution to the accumulated stress, most likely due to the detachment of In atoms from the WL entering either the liquid-like phase at the surface, or  the developing nanostructures, where the stress relief is more efficient. According to Bottomley, the In-Ga intermixing (Fig.~\ref{fig:Fig3}b) induces a partial melting of the surface and, consequently, a reduction of the accumulated stress, as we observe.\cite{bottomley1998} The cause of the surface melting is the lower binding energy of InAs vs. GaAs, and the process is assisted by the strain field gradients on the crystal surface. The same strain fields favour the migration of In atoms towards the nanostructures, in agreement with the analysis made by Joyce (Fig.~\ref{fig:Fig3}c).\cite{joyce2000} This interpretation also explains the results shown in (Fig.~\ref{fig:Fig2}b). As the number of In/Ga cycles increases, the strain gradients at the growth front become more intense. In such situation, a larger fraction of dissolved In atoms can more efficiently reach the QPs leading to a more pronounced reduction of the compressive stress. At the same time, since less In atoms remain in the liquid-like phase at the surface, the ulterior deposition of Ga also causes less stress accumulation as observed in the figure. 


\begin{figure}
	\centering
	\includegraphics{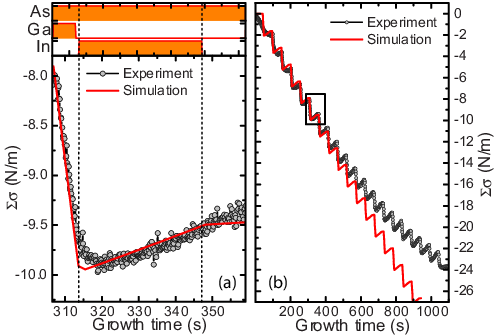}
	\caption{(a) One period and (b) full  QPs MOSS curve, including experiment and simulation.}
	\label{fig:Fig4}
\end{figure} 

We have simulated this scenario by numerical modelling of the strain fields using static continuum elasticity theory and a finite differences algorithm.\footnote{See Supplemental Material at [URL will be inserted by publisher] for further details about the simulation and modelling of the MOSS curve.} The strain fields are calculated at different stages by introducing the instantaneous compositions and morphologies. For each point, we integrate the calculated strain fields simulating an accumulated stress curve which can be compared with the experimental one. Our method is new and could be combined with more sophisticated and precise approaches, such as those presented by Shchukin \textit{et al}.,\cite{Shchukin2008}, to predict the shape, composition and temporal evolution of the nanostructures for a given static situation. Such effort is out the scope of the present work and therefore, to do the simulation, the basic information regarding the QPs size and composition has been attained by other experimental techniques or assumed. 

For simplicity, we consider that after several periods a steady state has been reached with a constant amount of In segregating from layer to layer. The presence of the seed is neglected, as it only affects the first periods of the QPs sequence. The QPs formation has been simulated by stacking lens shaped QDs, 24 nm wide, 2 nm high, and separated 3.5 ML ($\sim$1 nm) from each other. This implies an overlap of consecutive QD layers as required for the formation of the QPs.  A homogeneous InGaAs alloy is assumed for the QPs (63\% In).  At each Ga deposition step, the surrounding matrix is modelled as a 3.5 ML thick InGaAs layer with 24\% In. This indium linearly reduces to 17\% during the growth of the next InAs layer, accounting for the melting and surface diffusion process described above. We consider in the simulation a homogeneous InGaAs alloy, without distinction between WL and capping, as it has been reported.\cite{li2007} The In atoms detached from the matrix, together with the new incoming atoms, are used to fabricate the new QD or left in the liquid-like phase during the In deposition step, and no further shape or composition change is considered for the QDs during capping. The liquid-like phase has no effect on the stress, playing only a role of In reservoir. Each period of this structure (QD + capping) has an In content equivalent to 0.7 ML of InAs once buried, matching the nominal values used in the experiment.

The accumulated stress curve that results from this modelling is depicted in Fig.~\ref{fig:Fig4}a together with the experimental one. As it can be seen, even with this idealized sequence, the simulated curve exhibits the same oscillations than the experimental one. With a minimum set of assumptions, the simulated values reproduce the sign reversal of the $\Sigma\sigma$ curve observed during the In deposition step, strongly supporting the partial melting and migration effects discussed in previous experimental and theoretical works.\cite{Tersoff1998, Koduvely1999, joyce2000, bottomley2002} The overall agreement shown in Fig.~\ref{fig:Fig4}b is also remarkable although, in the long term the experimental curve accumulates less stress than the simulated one. This is reasonable since the strength of the inhomogeneous strain fields increases with each cycle and although in the simulation the fraction of In that incorporates to the QPs has been kept constant, in the experiment we expect an augmented surface migration efficiency and hence a reduced amount of liquid-like In, as it was discussed in view of the results shown in Fig.~\ref{fig:Fig2}b.  According to the simulation, we can conclude that at least 30\% of the In contained in the matrix surrounding the QPs is drained into them during the InAs deposition.

In summary, we have measured \textit{in-situ} and in real time the stress accumulated during the growth of InAs/GaAs quantum posts and developed a novel numerical method to analyze its temporal evolution. We conclude that In atoms detach from the InGaAs crystal surface during In deposition, and migrate towards the nanostructures in the presence of inhomogeneous strain fields, causing a temporal reduction of accumulated stress during In deposition. Our work lends experimental support to previous theoretical proposals on the subject of group III metal diffusion on III-V crystal surfaces, providing new experimental data  not easily attainable by other techniques. Furthermore, the methodology developed is not exclusive of quantum posts and can be combined with other established methods to get further insight in the growth kinetics of quantum nanostructures and other strained heteroepitaxial systems.  

We  acknowledge  the  financial  support by Spanish MINECO thorough grants ENE2009-14481-C02-02 and TEC2011-29120-C05-04, and by Spanish CAM thorough grants S2009/ESP-1503, S2009/ENE-1477 

%

\end{document}